\documentclass[aps,prd,twocolumn,groupedaddress]{revtex4-1}
\usepackage{graphicx}
\usepackage{amsmath}
\usepackage{amsfonts}
\usepackage{inputenc}
\usepackage{epsfig}

\begin{document}

\title{Application of Graphene within Optoelectronic Devices and Transistors}

\author{F.V. Kusmartsev}
\affiliation{Department of Physics, Loughborough University, Loughborough, Leicestershire, LE11 3TU, United Kingdom}

\author{W.M. Wu}
\affiliation{Department of Physics, Loughborough University, Loughborough, Leicestershire, LE11 3TU, United Kingdom}

\author{M.P. Pierpoint}
\affiliation{Department of Physics, Loughborough University, Loughborough, Leicestershire, LE11 3TU, United Kingdom}

\author{K.C. Yung}
\affiliation{Department of Industrial and Systems Engineering, The Hong Kong Polytechnic University, Hung Hom, Kowloon, Hong Kong, China}

\date{\today}

\begin{abstract}
Scientists  are  always  yearning   for  new and  exciting  ways  to  unlock  graphene's true  potential. However, recent reports suggest this two-dimensional material may  harbor some unique properties, making  it  a  viable  candidate  for  use  in  optoelectronic  and  semiconducting  devices.   Whereas on one hand, graphene is highly transparent due to its atomic thickness, the material does exhibit a strong interaction with photons.  This has clear advantages over existing materials used in photonic devices such as Indium-based compounds. Moreover,  the  material  can  be  used  to  'trap' light and  alter  the  incident  wavelength,  forming  the  basis  of the  plasmonic  devices.   We also  highlight upon graphene's nonlinear  optical  response  to  an  applied  electric  field,  and  the  phenomenon of saturable absorption. Within the context of logical devices, graphene has no discernible band-gap. Therefore, generating one will be of utmost  importance.   Amongst  many   others,  some  existing methods  to open this band-gap  include  chemical  doping, deformation  of the honeycomb  structure, or  the  use  of  carbon  nanotubes   (CNTs).    We  shall  also  discuss  various   designs  of  transistors, including  those which incorporate  CNTs, and others which exploit  the idea of quantum  tunneling. A  key  advantage  of  the  CNT transistor  is  that   ballistic transport   occurs  throughout   the  CNT channel,  with short  channel  effects  being minimized.  We shall  also discuss  recent  developments  of the  graphene  tunneling  transistor, with  emphasis  being  placed  upon  its  operational   mechanism. Finally, we provide  perspective   for  incorporating  graphene  within  high  frequency  devices,  which do not require  a pre-defined  band-gap.
\end{abstract}

\pacs{}

\maketitle

\section{Introduction}

Two-dimensional materials have always been considered  unstable due to their thermal fluctuations \cite{opto1,opto2}, in  what were famously referred to  as the  Landau-Peierls arguments. However, many scientists have not given up hope that such two-dimensional structures exist. In 2004, a research team based in Manchester successfully segregated graphene flakes from a graphite sample via `mechanical exfoliation' (more commonly referred to as the scotch-tape method) \cite{opto1,opto3,opto4,opto5,opto6}.  They witnessed a full preservation of graphene's hexagonal honeycomb structure, with astounding electrical, thermal and optical characteristics.

\begin{figure}[h]
	\centering
		\includegraphics[width=0.5\textwidth]{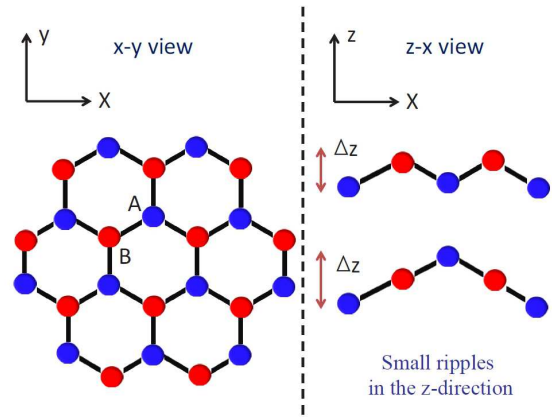}
	\caption{The honeycomb structure of graphene is presented in the left-hand figure. The right-hand figure depicts small quantum corrugations of the sub-lattices A and B, which are shifted in the transverse direction by a small fraction of the inter-atomic spacing `a'.}
	\label{optofig1}
\end{figure}

Graphene  is an  allotrope of carbon -  other examples include diamond, fullerene and charcoal, all with their own unique properties. Usually graphene will be found in the form of highly ordered pyrolytic graphite (HOPG), whereby individual graphene layers stack on top of one another to form a crystalline lattice.  Its stability is due to a tightly packed, periodic array of carbon atoms \cite{opto7} (cf.  FIG. \ref{optofig1}), and an $sp^{2}$  orbital hybridization - a combination of orbitals $p_{x}$  and $p_{y}$  that constitute the $\sigma$-bond.  The final $p_{z}$  electron makes up the $\pi$-bond, and is key to the half-filled band which permits free-moving electrons \cite{opto8}. In total, graphene has three $\sigma$-bonds and one $\pi$-bond. The right-hand portion of FIG. \ref{optofig1}, emphasizes how small displacements of the sub-lattices A and B can be shifted in the z-direction \cite{opto9}.

Moreover, graphene's mode of preparation will have a strong influence upon its overall quality and characteristics. As conducted by Geim et al. [10], mechanical exfoliation consists of gradually stripping more and more layers from a graphite sheet, until what remains are a few layers of graphene. In terms of overall mobility and the absence of structural defects, this method will produce the highest quality material. Other methods such as vacuum epitaxial growth or chemical vapour deposition (CVD), each have their own merits, but will generally lead to  inferior quality.    For  a  more in-depth discussion of the  available manufacturing methods, one can refer to \cite{opto1,opto11,opto12,opto13,opto14,opto15,opto16,opto17,opto18,opto19,opto20,opto21}.

On the other hand, graphene is highly impermeable  \cite{opto7} - the mobility can become severely compromised upon molecular attachment.  Yet, this apparent flaw has immediate applications for molecular sensors.  By  monitoring the deviation of electrical resistivity \cite{opto22,opto23}, one could, for example, envision novel smoke detection systems.  So too is graphene more than 100 times stronger than steel \cite{opto7}, possessing a Young's modulus as large as 1TPa  \cite{opto24}. Together with its outstanding electrical \cite{opto8}, thermal \cite{opto25,opto26,opto27} and in particular, optical properties \cite{opto11,opto24,opto28,opto29}, graphene has thus become a widely sought after material for use in future semiconducting and optoelectronic devices \cite{opto12,opto30}.\\ \\ \textbf{Electrical  Mobility} - As a material, graphene harbors some remarkable qualities; highly elastic due to its monolayer structure, and more conductive than copper with mobilities reaching up to $200,000$cm$^{-2}$V$^{-1}$s$^{-1}$ for perfect structures \cite{opto24,opto31,opto32}.  Charge  carriers in graphene travel with a Fermi velocity $v_{F} =\sqrt{3}\gamma_{0}a/2\hbar\approx 10^{6}$ms$^{-1}$.  Here, $\gamma_{0}\approx 3$ eV  is the energy required to 'hop' from one carbon atom to the nearest neighbor, $a\approx 1.42${\AA}  is the inter-atomic spacing between two neighboring carbon atoms, and $\hbar$ is Planck's constant \cite{opto8,opto24,opto33}. This Fermi velocity is approximately 1/300 the speed of light, thus presenting a miniaturized platform upon which to test many features of quantum electrodynamics (QED) \cite{opto1}.  Theoretical studies with graphene show that  the density of states (DoS)  of electrons approaches  zero at the Dirac point.  However, a minimum conductivity $\sigma_{0}\approx4e^{2}/h$ has been displayed \cite{opto1}, which is approximately double that for the conductance quantum \cite{opto34,opto35}. Even at room temperature, electrons can undergo long range transport with minimal scattering \cite{opto1,opto32,opto36}.\\
\\ \textbf{Thermal   Conductivity}  -  Heat  flow in suspended graphene was recently shown to  be mediated by ballistic phonons, and has been verified by Pumarol et al \cite{opto25} with the use of high resolution vacuum scanning thermal microscopy.  However, when considering multiple layers of graphene, this transport will be reduced due to an increase of inelastic scattering. The same is observed for graphene coated upon a substrate - the mean free path of thermal phonons degrading to less than 100\,nm.  Nevertheless, graphene on a silicon substrate can still retain a thermal conductivity of around 600\,Wm$^{-1}$K$^{-1}$  \cite{opto26} - even higher than copper. Whilst  the mechanism of heat transport across the graphene-substrate  interface remains unknown  \cite{opto37}, it is possible this may be linked to the in-plane thermal conductivity \cite{opto25,opto27}.\\ \\ \textbf{Optical  Response} - Graphene's atomic thickness makes it almost perfectly transparent to visible light \cite{opto9,opto38}, allowing such a material to become widely accessible to a number of applications. These cover everything from photovoltaic cells, to graphene photonic transistors \cite{opto30,opto33,opto39,opto40}. Being a single layer of carbon atoms, graphene also exhibits many interesting photonic properties. As  such, our focus will be directed mainly upon those which are associated with applications to optoelectronic devices. The transmittance between multiple graphene layers, how optical frequency relates to conductivity,  nonlinear optical response, saturable absorption and plasmonics will all be discussed in later sections.

Most semiconducting photonic devices will be governed in some way by electron excitation and electron-hole recombination. Excitation  refers to an electron absorbing photon energy of a very specific wavelength within the allowed energy bands.  On  the other hand, recombination is a process which leads to the emission of photons (cf.  electro-luminescence) \cite{opto38}. Gallium  arsenide (GaAs),   indium functional compounds and silicon are all common semiconductors for use in photonic devices \cite{opto41,opto42}.  However, graphene exhibits a strong interaction with photons, with the potential for direct band-gap creation and thus being a good candidate for optoelectronic and nanophotonic devices \cite{opto43}. Its strong interaction with light arises due to the Van Hove singularity \cite{opto44}. Graphene also possesses different time scales in response to photon interaction, ranging from femtoseconds  (ultra-fast) to picoseconds \cite{opto43,opto45}. Overall,  graphene could easily be an ideal candidate for transparent films, touch screens and light emitting cells.  It may even be used as a plasmonic device which confines light, and altering the incident wavelength. We shall elaborate upon this in later sections.

\section{Energy Spectrum, Band-Gap and Quantum Effects}

\begin{figure}[h]
	\centering
		\includegraphics[width=0.4\textwidth]{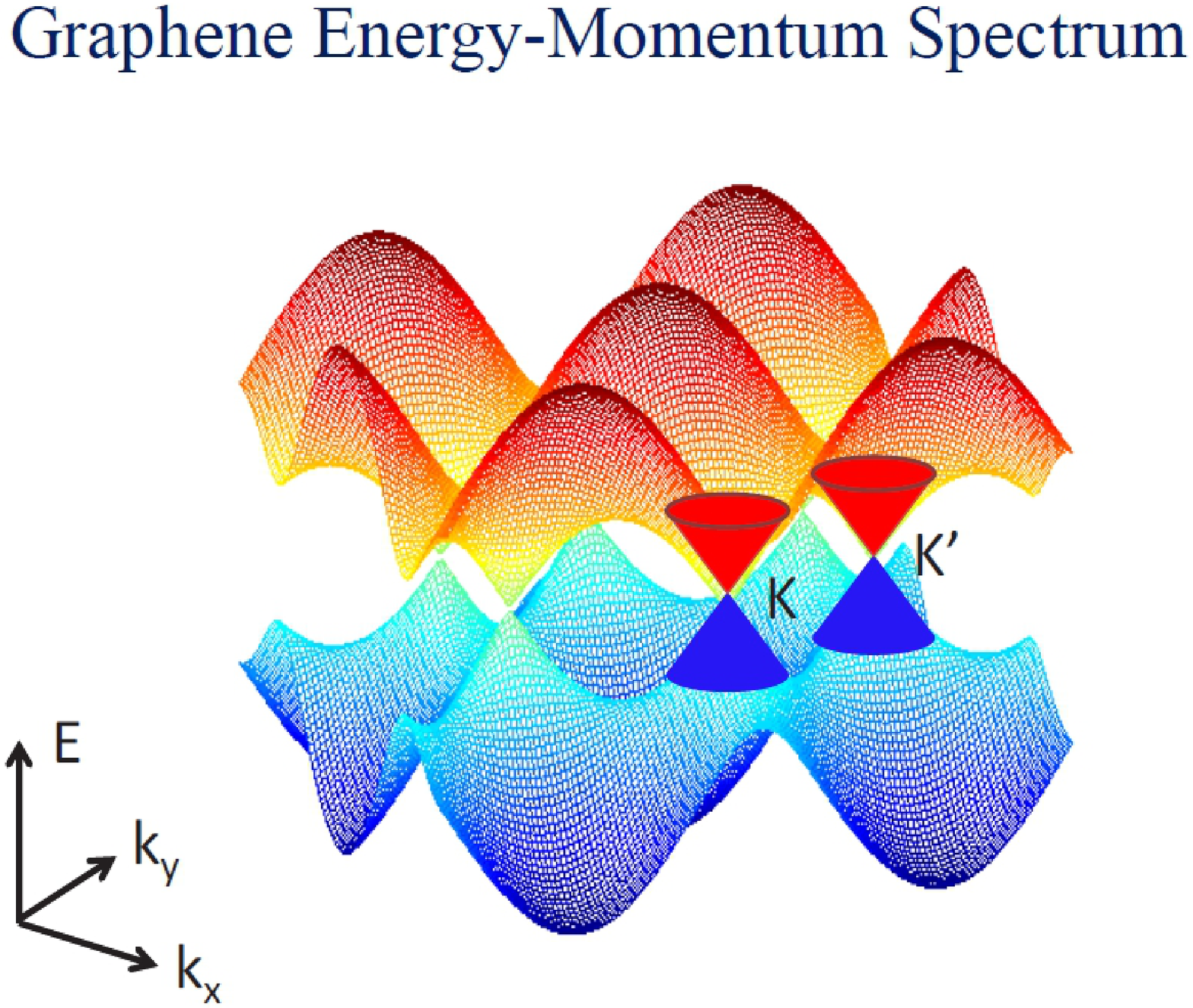}
	\caption{The energy-dispersion spectrum as given by Eq.(\ref{optoeq1}). Here, the $z$-axis represents the energy $E({\rm {\bf k}})$, with the $x-y$ plane corresponding to the momentum {\bf k}=$({\rm k_{x},k_{y}})$. Dirac cones are located at both the K and K' points of the Brillouin zone.}
	\label{optofig2}
\end{figure}

Theoretical studies of monolayer graphite (i.e.,  graphene) first began in 1947 by Wallace \cite{opto8}, who considered a simple tight-binding model with a single hopping integral. This model takes into account the hopping of an electron from one carbon atom to its first and second nearest neighbors only.  Wallace's conclusions were stark; an electrical conductivity should theoretically  exist for two-dimensional graphene. To elaborate; at six positions of the Brillouin zone, Dirac points (K  and K')  exist. These are points in momentum space for which the energy $E(\textbf{\rm{p}}_{0}) = 0$, where $\textbf{\rm{p}}_{0}$  = $\hbar$\textbf{K} (or $\hbar$\textbf{K}').  Here, we have denoted the momentum as a vector \textbf{p} = ($\rm{p_{x} , p_{y}}$) = $\hbar$\textbf{k}, where \textbf{k} = ($\rm{k_{x}, k_{y}}$) is the wave vector \cite{opto1}. The energy eigenvalues were found to take a gapless form \cite{opto8},

\begin{equation}
E^{\pm}(\rm{k_{x},k_{y}})=\pm\gamma_{0}\sqrt{1+4\cos\frac{\sqrt{3}\rm{k_{x}}a}{2}\cos\frac{\rm{k_{y}}a}{2}+4\cos^{2}\frac{\rm{k_{y}}a}{2}}
\label{optoeq1}
\end{equation}\\
where the plus and minus signs refer to the upper and lower half-filled bands respectively \cite{opto24,opto34}. By expanding the above equation in the vicinity of the K or K'  points, one can thus obtain a linear dispersion relation that is given by $E^{\pm} = \pm v_{F} \hbar |\delta \textbf{k}|$, where \textbf{k} = \textbf{K}+$\delta$\textbf{k}.  These constitute what are known as Dirac cones, and are clearly emphasized by FIG. \ref{optofig2}.  Here, a direct contact of the conduction and valence bands is found \cite{opto5,opto8,opto10,opto11}, thus pertaining to a zero energy band-gap $E_{g}$ \cite{opto1,opto8,opto12}. Therefore, generating a band-gap in graphene will be essential for its application within semiconducting devices (e.g.,  transistors). On the other hand, graphene may secure its place in high-frequency devices, which do not require a logical OFF state \cite{opto38}.

\begin{figure}[h]
	\centering
		\includegraphics[width=0.45\textwidth]{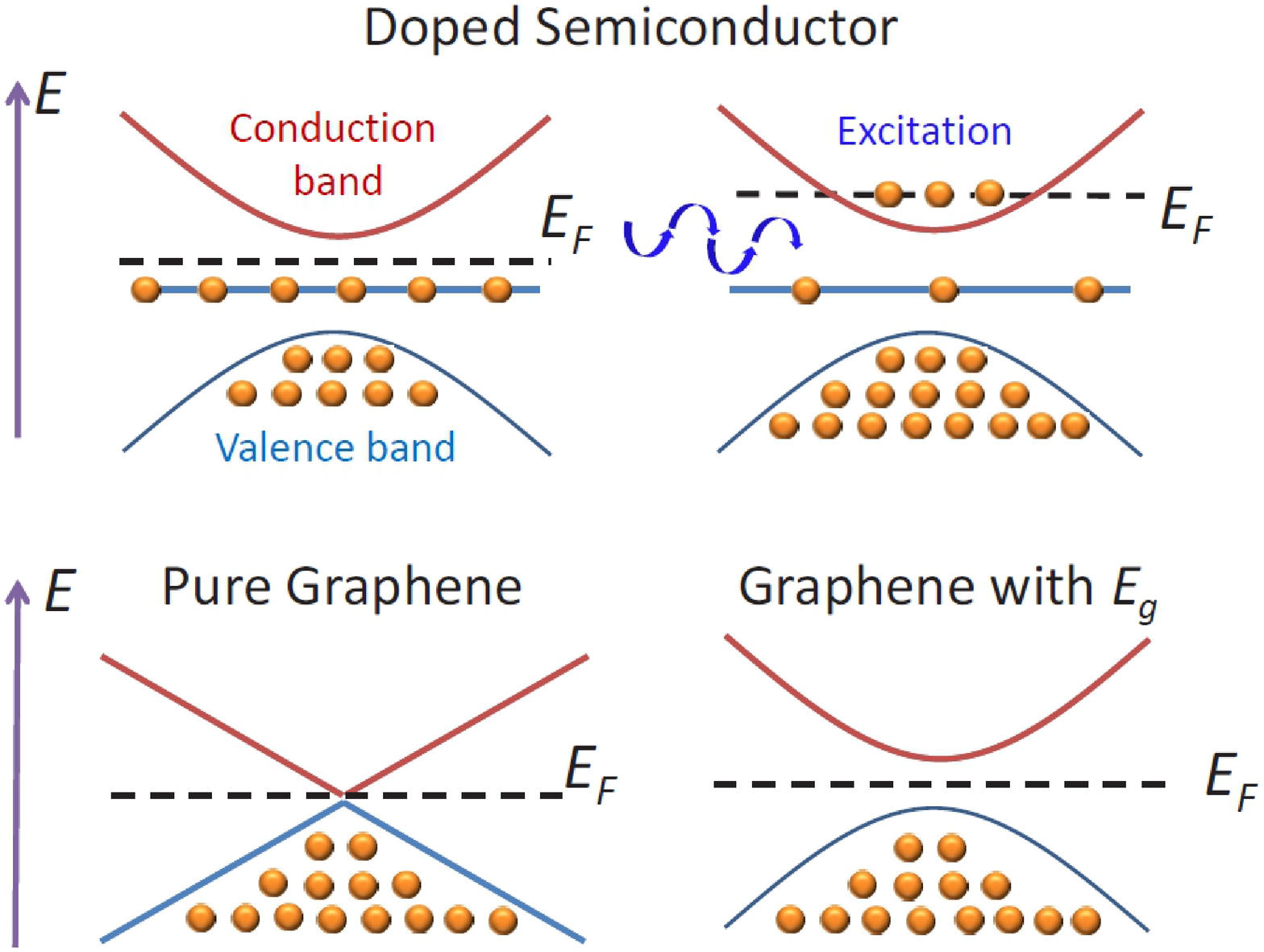}
	\caption{The upper half of this figure depicts the electronic band structure of a doped semiconductor. Typically, the band-gap for a doped semiconductor is very small, with only a small energy being required to excite an electron from the valence to conduction band. The lower figure shows the electronic band structure for graphene. For pure samples, no energy band-gap $E_{g}$ exists. In principle, an energy band-gap can be created via many methods.}
	\label{optofig3}
\end{figure}

\ \\ \textbf{A. 	Dirac Energy-Momentum Dispersion}\\ \\ 
Supposing we consider the Hamiltonian $\hat{H}$ as given by Wallace \cite{opto8}  - in the low energy limit, spinless carriers in graphene  possess a zero effective mass, and are well approximated by the relativistic Dirac Hamiltonian $\hat{H}$ \cite{opto34},

\begin{equation}
\hat{H}= v_{F}\hbar\hat{\sigma}\dot\delta \textbf{k}\ ,
\label{optoeq2}
\end{equation}\\
where $\hat{\sigma}\dot\delta\textbf{k}=\sigma_{x}\delta\rm{k_{x}}+\sigma_{y}\delta k_{y}$. Here, $\hat{\sigma} = (\sigma_{x},\sigma_{y})$ is the vector of $2\times2$ Pauli matrices:
\begin{equation}
\sigma_{x}=\bordermatrix{& & \cr
	& 0&1 \cr
	& 1&0 \cr}\ \ \ ,\ \ \ \sigma_{y}=\bordermatrix{& & \cr
	& 0&-i \cr
	& i&0 \cr}\ .\label{optoeq3}\end{equation}\\
The spinor wave function $\psi$ of graphene can be obtained from,

\begin{equation}
\hat{H}\psi=E\psi\ ,
\label{optoeq4}
\end{equation}\\
where $E$ denotes the energy eigenvalues of $\hat{H}$ \cite{opto24}. Here, $\psi = (\psi_{A} , \psi_{B})^{\rm{T}}$  is a vector containing the two component wave function.  These components represent the sub-lattices A and B accordingly \cite{opto34}.\\ \\
\textbf{B. 	Band-Gap Creation}\\ \\ 
Generally speaking, the electrical conductivity of a material can fall into one of three groups: conductors, insulators, semiconductors  \cite{opto46,opto47,opto48}. For a conductor, electrons are able to move freely in the conduction band since electron states are not fully occupied. However, the conduction and valence band may sometimes be separated by an energy band-gap $E_{g}$ (e.g.,  for insulators and semiconductors), thus preventing the free movement of electrons in the conduction band.  For an insulator, an electron requires a huge energy in order to excite from the valence to conduction band.  A small band-gap is present for semiconductors, with an electronic band structure that is parabolic in shape \cite{opto12,opto31}. Doped semiconductors will make the band-gap even smaller, and hence more easy to control (cf.  FIG. \ref{optofig3}).

Graphene's high mobility makes it a particularly enticing material for use in electronic devices. However, we have already mentioned that in the vicinity of the Dirac point, graphene possesses  a conical band structure which is gapless (i.e., $E_{g} = 0$) \cite{opto1,opto10}.  Thus our main concern with regards to logical devices, is the absence of this well-defined  OFF state pertaining to zero current flow. To rectify this, we must open up an energy band-gap such that $E_{g}\neq0$.  With  regards to optoelectronic devices, a tunable band-gap can specify the range of wavelengths which can be absorbed. The energy bands for pure graphene, and graphene with a small band-gap $E_{g}$ are displayed in FIG. \ref{optofig3}.  The Fermi energy level $E_{F}$   is situated at the Dirac point for pure graphene \cite{opto3}. For graphene that has been modified to include a band-gap, an energy is required to excite electrons from the valence to conduction band, and hence an ON/OFF state regime is established \cite{opto10}. Amongst many others, existing methods include the use of carbon nanotubes (CNTs), graphene nanoribbons (GNRs)  or even bilayer graphene \cite{opto1,opto31,opto38,opto49}.  However, it is important to note that  although bilayer graphene does possess  a zero energy band-gap, an applied electric field can be used to create one \cite{opto9,opto12,opto31}. Other methods include deformed structures, graphene oxide (GO) \cite{opto12,opto22,opto23}, and also the use of chemical doping via compounds such as Boron Nitride (BN) \cite{opto50,opto51}. The  idea here, is that  the doped atoms alter graphene's honeycomb structure, similar to deformation or localized defects \cite{opto52,opto53}. All in all, one has to note that the aforementioned methods are not well-developed  enough to maintain a high mobility.    Much  more exotic concepts are required, which we shall now discuss.\\ \\
\textbf{C. 	Quantum Phenomena}\\ \\ 
Of  its  more surprising attributes,  graphene has also displayed signs of anomalous quantum behaviors, even at  room temperature \cite{opto5,opto54}.  We  shall briefly discuss two key phenomena in particular.\\ \\
\textbf{Quantum Hall Effect:}  QHE   has been observed for both  single and  bilayer graphene \cite{opto11,opto55,opto56}, in the presence of a magnetic field B.  The Landau levels for graphene are given by,

\begin{equation}
E_{Landau}=\sqrt{|2e\hbar v_{F}^{2}\rm{B}|}\ ,
\label{optoeq5}
\end{equation}\\
where $e$ is the electric charge, and $j\in\mathbb{Z}$ is the Landau index \cite{opto5,opto54}. In conventional 2-D semiconductors, the Landau levels are $E = \hbar\omega_{c}(j + 1/2)$, where $\omega_{c}$ is the cyclotron frequency \cite{opto5,opto54}.  The anomalous energy spectrum for graphene subject to a B  field leads to a one half shift of the minimum conductivity at the zero energy Landau level, whereas traditional QHE  semiconductors give an integer one \cite{opto5,opto54}. The Hall conductivity $\sigma_{H}$  is therefore given by \cite{opto24,opto34}, 

\begin{equation}
\sigma_{H}=g\left(j+\frac{1}{2}\right)\frac{e^{2}}{h}\ ,
\label{optoeq6}
\end{equation}\\
where $g$ is the degeneracy.  For graphene, a fourfold degeneracy exists - two spins, and the valley degeneracy of the K  and K'  Dirac points \cite{opto5}. Additionally,  the fractional QHE  has been observed for both monolayer and bilayer graphene (cf.  for details \cite{opto5,opto55,opto56}).\\ \\ 
\textbf{Klein    Tunneling:}     Intuition   states  that   if   a   particle's  kinetic  energy  $KE$   is  less than some value $U$, then it  will be physically incapable of surpassing a potential barrier of  the  same energy $U$.   However, quantum  mechanics states that  a  particle is  able to tunnel the potential barrier $U$ with a certain decay probability \cite{opto5}. Furthermore, relativistic quantum mechanics permits a remarkable phenomenon called Klein tunneling. Much like a freight train instead taking a tunnel from one side of a mountain to the other, an electron can perform a similar process \cite{opto54,opto57}. This occurs when an electron experiences a strong repulsive force from the barrier $U$, and hence induces a hole inside the barrier \cite{opto5,opto55,opto56}. This  leads to  a  matching  of  the  energy spectrum inside and  outside the  barrier, with the transmission probability becoming very close to  one \cite{opto54}.  A  perfect transmission is demonstrated for square potentials only,  and is dependent upon the energy $KE$, and the angle of incidence $\theta$ relative to the barrier \cite{opto58}. Confined bound states will arise for energies close to the Dirac  point \cite{opto58}.  Further details regarding how this confinement effect may relate to the special waveguide geometry has been discussed in references \cite{opto59,opto60,opto61,opto62,opto63}.

\section{Photonic Properties}

Optical communication networks are ubiquitous nowadays, affecting our everyday lives. A fiber-optic cable provides a much wider bandwidth, and less energy loss than some traditional copper wiring \cite{opto41,opto42}.   According  to  the  Shannon-Hartley theorem \cite{opto64}, the  maximum capacity of a channel is given by

\begin{equation}
max(C)=B\log_{2}\left(1+\frac{P_{s}}{P_{n}}\right)\ ,
\label{optoeq7}
\end{equation}\\
where $B$ is the channel bandwidth, and $P_{s}$   and $P_{n}$  are the average signal and noise powers respectively.   It  is therefore obvious that  optical cable provides a much larger channel capacity, where $P_{s}/P_{n}\gg1$.

\begin{figure}[h]
	\centering
		\includegraphics[width=0.4\textwidth]{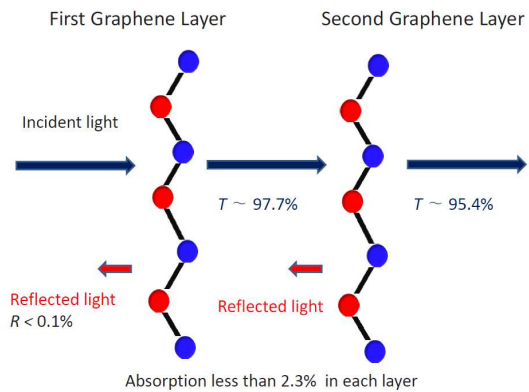}
	\caption{Incident light passes through the to layers of graphene. The transmission, absorption and reflection coefficients are all shown. Each layer of graphene only absorbs 2.3\% of incident light, transmitting around 97.7\%, and reflecting less than 0.1\%.}
	\label{optofig4}
\end{figure}

When optical and electronic devices work together (e.g.,  a modulator), light signals are converted into an electrical equivalent. Generally speaking, the term 'optoelectronic' refers to an optical (photonic) electronic device, which transmits signals via light waves, or electron-photon interaction \cite{opto41,opto42}. A photonic device can be made of semiconductors, either being integrated into electronic circuits or transistors. Optoelectronics also play an important role as the mediator of optical communication.  Devices will typically operate with an optical frequency ranging from ultraviolet to infrared (400--700\,nm) \cite{opto41,opto42}, although graphene photonic devices can possess an even wider spectrum than this \cite{opto65}.\\ \\
\textbf{A. 	Transmittance Properties}\\ \\ 
As emphasized by FIG. \ref{optofig4}, a single layer of graphene absorbs a mere 2.3\% of incident light, allowing around 97.7\% to  pass through.   Wavelengths typically  range from the  infrared to  ultraviolet regions \cite{opto33}.  The  transmittance $T$  of single-layer graphene (SLG) can be approximated by the following Talyor expansion \cite{opto33,opto39,opto40,opto66,opto67}

\begin{equation}
T = \frac{1}{(1+\alpha\pi/2)^{2}}\approx1-\alpha\pi\approx97.7\%\ ,
\label{optoeq8}
\end{equation}\\ 
where $\alpha = e^{2}/c\hbar\approx 1/137$  is the fine structure constant.  For multiple layers of graphene, this can be roughly estimated by

\begin{equation}
T\approx(1-N\alpha\pi)\ ,
\label{optoeq9}
\end{equation}\\
where $N$  is number of layers (cf.  Bao \textit{et al.} \cite{opto39}). For example, the transmittance of bilayer graphene ($N = 2$) is around 95.4\% (cf.  FIG. \ref{optofig4}). Indium Tin Oxide (ITO)  is a semiconductor which is typically used in photonic devices, with a transmittance of around 80\% \cite{opto33}. It is therefore obvious that graphene film has a clear advantage over ITO. Bonaccorso \textit{et al.} \cite{opto33} also point out that the resistance per unit area for ITO  is much smaller than for graphene. However, this value can be minimized by increasing the concentration of charge carriers via methods such as doping.

The degree of reflection from SLG is almost negligible, just less than 0.1\% \cite{opto33}. Avouris \textit{et al.} \cite{opto40,opto66,opto67} also mention that graphene shows a strong interaction of photons, much stronger than  some traditional  photonic materials per unit  depth.    It  is also surprising that absorption can rise from 2.3\% to around 40\% with high concentration doping \cite{opto40,opto66,opto67}. Unquestionably,  these properties present graphene as an excellent candidate for use in photonic applications.\\ \\
\textbf{B. 	Optical Conductivity}\\ \\
As mentioned by Avouris \textit{et al.}  \cite{opto40,opto66,opto67}, graphene possesses a universal optical conductance  $G_{op}  = e^{2}/4\hbar$.  In general, the optical conductivity $\sigma_{op}$  depends upon the frequency $\omega$, Fermi energy $E_{F}$   (via chemical doping or an applied gate voltage), and the transition rate $\Gamma$.  Moreover, the optical conductivity can be divided into real and imaginary components, $\Re(\sigma_{op})$ and $\Im(\sigma_{op})$ \cite{opto33,opto39,opto40,opto66,opto67},

\begin{equation}
\sigma_{op}(\omega,E_{F} ,\Gamma) = \Re(\sigma_{op}) + i\Im(\sigma_{op})\ ,
\label{optoeq10}
\end{equation}\\
with energy loss originating from the imaginary part \cite{opto40,opto66,opto67}.

\begin{figure}[h]
	\centering
		\includegraphics[width=0.4\textwidth]{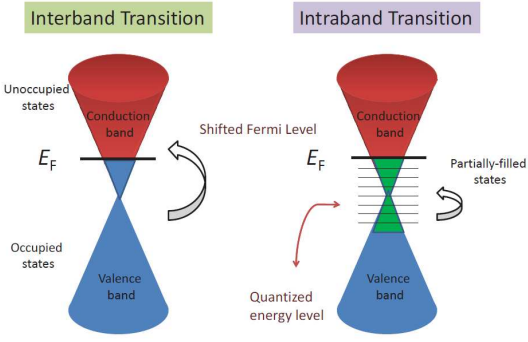}
	\caption{The Fermi energy level $E_{F}$ can be shifted upward due to either chemical doping or an applied electric field. Interband transitions refer to an electron `jumping' from the valence to conduction bands, satisfying the relationship $\hbar\omega\geq2E_{F}$. Intraband transitions refer to an electron moving through quantized energy levels, and requires less energy $(\hbar\omega<2E_{F})$, provided the states are not already occupied.}
	\label{optofig5}
\end{figure}

Bao \textit{et al.} \cite{opto39} further explain that the interband and intraband carriers' transitions are the major factors governing the optical conductivity $\sigma_{op}$ (cf.  FIG. \ref{optofig5}). Interband transitions refer to an exchange of charge carriers between the conduction and valence bands, whereas intraband transitions refer to a 'jump'  between quantized energy levels.  For carriers performing an interband transition (at high frequency), the energy of a photon $\hbar\omega$ should be satisfying the relationship $\hbar\omega\geq2E_{F}$ \cite{opto39}. For the low frequency THz range $(\hbar\omega < 2E_{F} )$, the intraband transition would be a significant contribution to the optical conductivity,  while interband transitions are prohibited in this range due to the Pauli exclusion principle (Pauli block) \cite{opto67}. It is important to note that a change in doping concentration would alter the Fermi energy $E_{F}$ , and hence the optical conductivity.   Bao \textit{et al.}  \cite{opto39}  state that  one can tune the optical conductivity by controlling the chemical doping (shift of $E_{F}$) and the frequency response. However, one must remain aware that  a high doping concentration may deteriorate the transmittance $T$ of graphene itself.

\begin{figure}[h]
	\centering
		\includegraphics[width=0.4\textwidth]{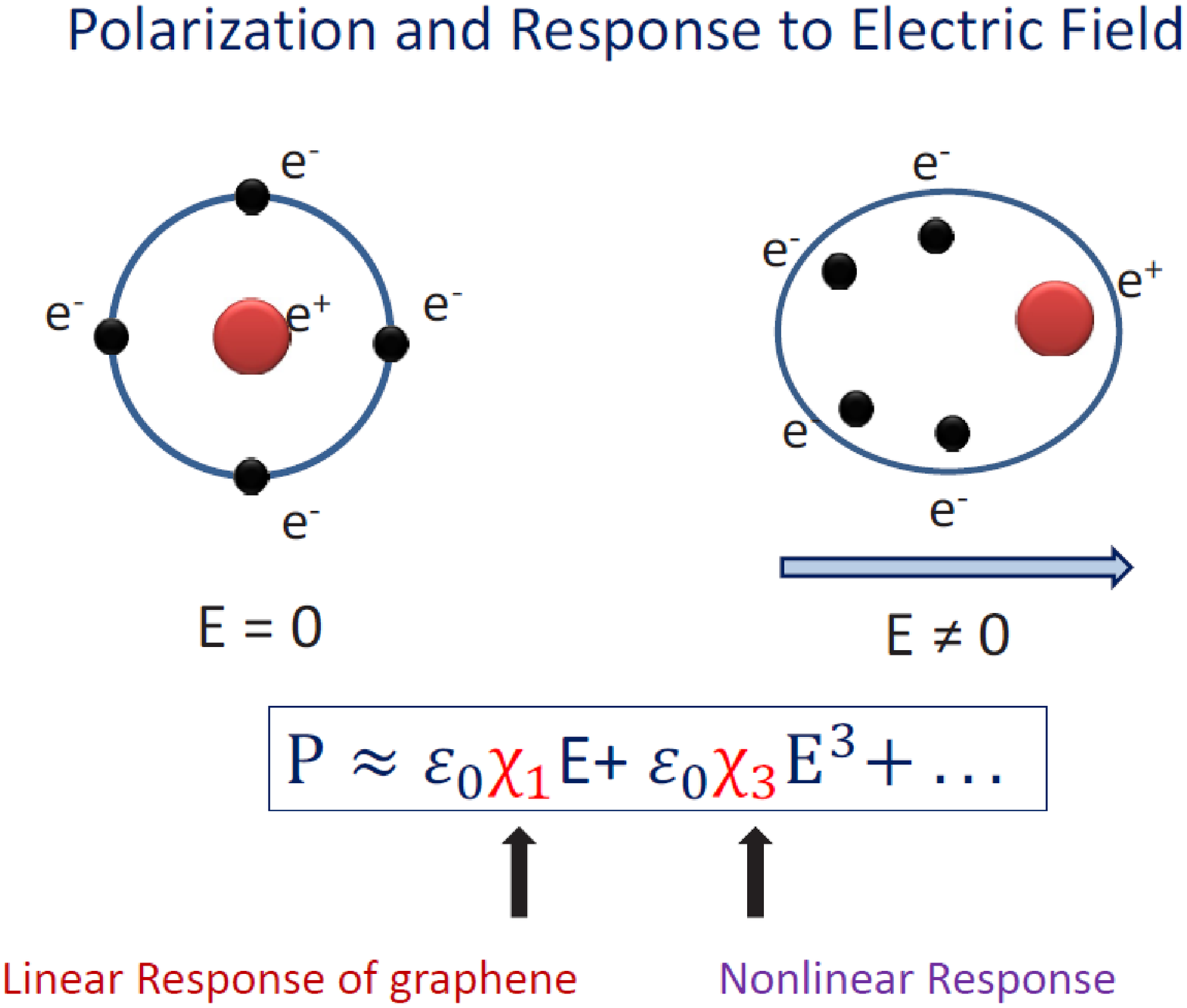}
	\caption{This figure shows the linear and nonlinear responses to an electric field E. The lefthand figure schematically represents an atom without electric field, whilst the right-hand figure has a non-zero electric field. The equation is related to the polarization response P. The linear susceptibility $\chi_{1}$ is usually associated with the refractive index, whereas the nonlinear susceptibility $\chi_{3}$ provides a unique contribution to the optical properties of graphene. Due to the symmetry of graphene’s honeycomb structure, the $\chi_{2}$ component is very small, and is therefore neglected here.}
	\label{optofig6}
\end{figure}

\ \\ \textbf{C. 	Linear  and  Nonlinear Optical Response}\\ \\
Graphene also exhibits a strong nonlinear optical response to an electric field, and is an important factor in modifying the shape of the wavefront for incident light \cite{opto40,opto66,opto67}. The displacement field D$_{z}$ is given by the dielectric response of an applied electric field E$_{z}$  along the 'z' direction, with polarization P(E$_{z}$) (cf.  FIG. \ref{optofig6});

\begin{equation}
\rm{D_{z}}  = \epsilon_{0}\epsilon_{r}\rm{E_{z}}  = \epsilon_{0}\rm{E_{z}}  + \rm{P(E_{z})}\ .
\label{optoeq11}
\end{equation}\\
Here, $\epsilon_{0}$  is the electric permittivity of free space, and $\epsilon_{r}$ is the relative permittivity.   The polarization response can be written in terms of a power series (cf.  for details \cite{opto39,opto68})

\begin{equation}
\rm{P(E_{z})}=C_{0}+\epsilon_{0}\sum_{j=1}^{\infty}\chi_{j}\rm{(E_{z})}^{j}\ ,                                 
\label{optoeq12}
\end{equation}\\
where $C_{0}$  is a constant associated with the hysteresis  (typically $C_{0}  = 0$), $\chi_{j}$  refers to the dielectric susceptibility of the j-th  order correction, and $\rm{(Ez)^{j}}$ is the j-th  power of $\rm{E_{z}}$. The linear dielectric susceptibility $\chi_{1}$ can again be divided into a real part $\chi_{R1}$ and an imaginary part $\chi_{I1}$ \cite{opto39}. The relative dielectric constant can then be expressed in terms of $\epsilon_{r}  = \chi_{R1} + 1$, with an optical refractive index $n_{op}$ given by,

\begin{equation}
n_{op}\approx\sqrt{\epsilon_{r}}=\sqrt{\chi_{R1}+1}\ .
\label{optoeq13}
\end{equation}\\                                            
Thus, the refractive index is determined by the real part of the linear susceptibility $\chi_{R1}$, as mentioned by Bao et al. \cite{opto39}. Meanwhile, the imaginary part of the linear susceptibility $\chi_{I1}$   corresponds  to  the  tangent  loss arising at  optical frequencies.  Bao  et  al.   \cite{opto39} also come to the conclusion that the second  order susceptibility $\chi_{2}$ is generally small, provided that the symmetry of the graphene honeycomb structure is not broken (i.e.,  flat).   The major contribution to the nonlinear response of graphene originates via the third  order term $\epsilon_{0}\chi_{3}{\rm E}_{z}^{3}$, which modifies the current density in graphene (cf.  for details \cite{opto39}).

\begin{figure}[h]
	\centering
		\includegraphics[width=0.4\textwidth]{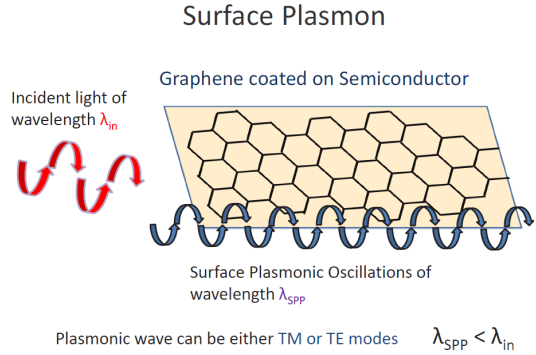}
	\caption{This figure shows the surface plasmonic wave for graphene coated on a semiconductor. Plasmonic waves are trapped, and oscillate along the graphene and semiconductor interface. Typically, the surface plasmonic wavelength $\lambda_{SPP}$ will be suppressed, and is much smaller than the incident wave $\lambda_{in}$. Either TM or TE wave modes can propagate along the plasmonic surface, depending upon the imaginary component of the optical conductivity.}
	\label{optofig7}
\end{figure}

\ \\
\textbf{D. 	Surface Plasmons}\\ \\
Surface plasmons describe a set of quantized charge oscillations of electrons and holes, acting upon the graphene-substrate interface \cite{opto39}  (cf.   FIG. \ref{optofig7}).   Plasmons, in general, interact with photons or phonons to form the surface plasmon polariton (SPP). At  present, aluminium, silver and gold are all ideal materials for plasmonic platforms \cite{opto39,opto30,opto66,opto67}. The basic idea is as follows - a dielectric material can be coated upon a graphene layer. Electrons then oscillate on the graphene-substrate interface, excited by the phonon or photon interactions of electromagnetic (EM)  fields \cite{opto39}. The SPP  wavelength $\lambda_{SPP}$ is normally suppressed, and much smaller than the incident wavelength $\lambda_{in}$ - the ratio of these wavelengths typically being around $\lambda_{in}/\lambda_{SPP}\approx 10 - 100$ \cite{opto39,opto40,opto66,opto67}. The plasmonic frequency $\omega_{SPP}$ on the graphene surface is proportional to the square root of the Fermi energy,  as given by \cite{opto33,opto39,opto40}

\begin{equation}
\omega_{SPP}\propto\sqrt{E_{F}}\propto n^{\frac{1}{4}}\ ,
\label{optoeq14}
\end{equation}\\
where $n$  is a  carrier density.    In  practice,  graphene can trap  incident light,  and an EM wave can propagate along the graphene surface in the THz  to infrared range \cite{opto40,opto66,opto67}. As  mentioned by Avouris et al. \cite{opto40,opto66,opto67}, the distance traveled for a plasmonic wave in graphene is around $d_{SPP}\approx 10-100 \lambda_{SPP}$.  Graphene is thus a suitable material for a waveguide. Bao et al. \cite{opto39} further remark that graphene is suitable for guiding transverse magnetic (TM)  waves when the imaginary part of the optical conductivity $\Im(\sigma_{op}) > 0$, and suitable for guiding transverse electric (TE)  waves when $\Im(\sigma_{op}) < 0$.\\ \\
\textbf{E. 	Saturable Absorption and  Optical Excitation}\\ \\
There is an interesting property which prevents graphene from absorbing photons at high intensity, and can be used to adjust the wavefront of the light \cite{opto40,opto66,opto67}. This is referred to as `saturable absorption', and is dependent upon the wavelength and incident light intensity. This will be elaborated upon in later sections when we discuss the saturable absorber and photonic (optical) limiter.

\begin{figure}[h]
	\centering
		\includegraphics[width=0.4\textwidth]{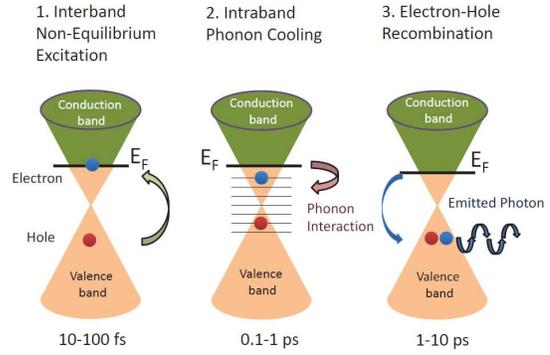}
	\caption{There are three time scales associated with optical response. The left-hand figure represents the interband non-equilibrium excitation, and lasts around 10-100\,fs. The middle figure relates to phonon cooling via the intraband interaction (0.1-1\,ps). Finally, the right-hand figure is the process of electron-hole recombination (1-10\,ps).}
	\label{optofig8}
\end{figure}

\begin{figure}[h]
	\centering
		\includegraphics[width=0.4\textwidth]{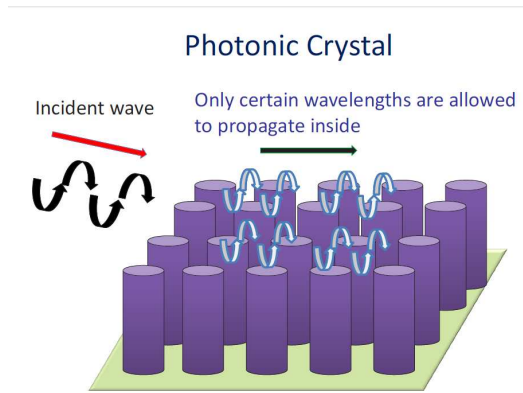}
	\caption{Here, we present a periodic photonic crystal lattice on the substrate. The lattice forms a band-gap, allowing only certain wavelengths to propagate inside. The photonic properties can therefore be controlled.}
	\label{optofig9}
\end{figure}

The timescale of graphene's response to the interaction of photons, phonons and electron-hole recombination can be divided into three regimes \cite{opto33,opto39,opto40,opto66,opto67} (cf. FIG. \ref{optofig8}). Graphene has a very quick response to incident photons, around 10-100\,fs, whereby 'hot' electrons are excited from the valence to conduction band \cite{opto33,opto39}.  This  also links to an excitation of the non-equilibrium state.   Electrons may then cool down via the intraband phonon emission, with timescales of 0.1\,ps \cite{opto40,opto66,opto67}. Finally, an electron-hole pair may recombine, thus emitting photons, and an equilibrium state being reached.  This  process takes a mere 1-10\,ps. It is important to note that these excitations and scattering processes are influenced by both topological defects of the lattices (e.g.,  dislocation and disclination) and boundary characteristics \cite{opto40,opto66,opto67}.\\ \\ 
\textbf{F. 	Graphene Photonic Crystal}\\ \\ 
The  photonic crystal is a kind of optical device, whereby a lattice can be periodically allocated upon or within a semiconductor \cite{opto69,opto70} (cf.  FIG \ref{optofig9}). A band-gap can be obtained in these periodic structures, and only a certain range of photon energies (i.e.,  frequencies) are allowed to propagate within.  The basic idea is that  the periodic dielectric behaves as a superlattice, with restriction being placed upon the wave properties of the electrons \cite{opto70}. Moktadir et al. \cite{opto71} find that  the graphene photonic crystal provides a wide transmission range, which can be tuned via an applied gate voltage.  It  has also been reported by Majumdar et al. \cite{opto71} that the resonance reflectivity can be increased fourfold via a slight 2\,nm shifting of the graphene crystalline structure (i.e.  dislocations). Graphene's flexible nature therefore offers numerous  applications.

\section{Graphene Optoelectronic Devices}

In this section, we will present ideas for optical devices which incorporate graphene, with emphasis being placed upon their photonic properties. The various photodetectors, optical modulator, and the photonic limiter (mode-locked laser) will all be discussed.\\ \\
\textbf{A. 	Photodiode and  Graphene Photodetector}\\ \\ 
The n-p or p-n junctions are comprised of two different semiconductors (n-type and p-type).  Electrons from the n-type semiconductor will flow across the p-type, whereas holes in the p-type will move to the n-type \cite{opto73}. In any case, a depletion layer is formed at the junction interface. In principle, n-p or p-n junctions can be forward or reverse biased \cite{opto73}. Since a  band-gap can be created in graphene (cf.   Section II.B),  it  is therefore feasible to conceive of a graphene-semiconductor junction.

\begin{figure}[h]
	\centering
		\includegraphics[width=0.4\textwidth]{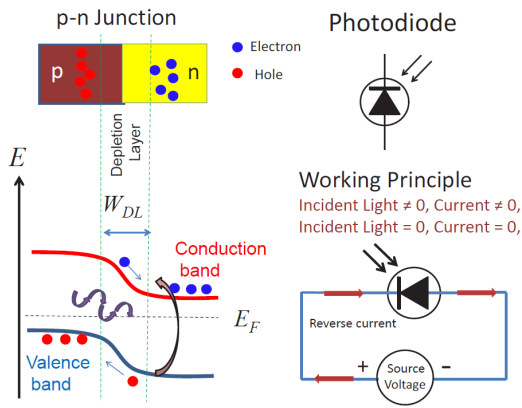}
	\caption{A schematic of the photodiode is shown. The basic idea is that a reverse current flows upon illumination of the photodiode. The figure also emphasizes that a depletion layer is formed at the interface of the p-n junction.}
	\label{optofig10}
\end{figure}

Photodiodes are a key component for use in logical devices.  It  is a current generating device that  is sensitive to incoming light (cf.  FIG. \ref{optofig10} for the p-n junction configuration). In the absence of light, the device carries a high resistance. However, incoming photons can break down some of the bonding within the compounds at the depletion layer (cf.  FIG. \ref{optofig10}). Electrons and holes will then be created, and hence a drift current $I_{d}$ flows across the diode
\cite{opto73}

\begin{equation}
I_{d}\approx c_{1} (1 - \exp(-c_{2}W_{DL} ))\ .
\label{optoeq15}
\end{equation}\\
Here, $c_{1}$  is a constant associated with electric charge and photonic flux, $W_{DL}$  is the width of the depletion layer, and $c_{2}$ is a constant associated with the photon energy and band-gap \cite{opto45}.

\begin{figure}[h]
	\centering
		\includegraphics[width=0.4\textwidth]{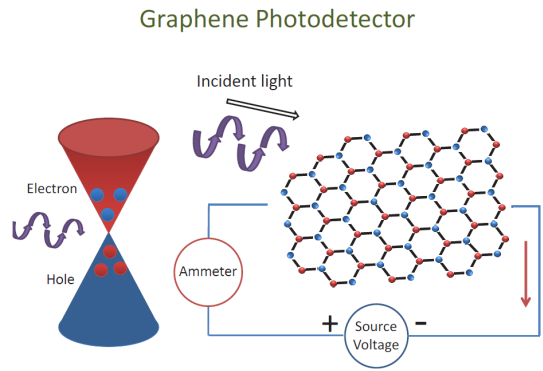}
	\caption{This figure shows how electrons in the valence band can be excited to conduction band by incident photons. The conductivity of graphene increases, and a measurable current can flow around the circuit. In practice, the idea can be used to measure incident light intensity, for example.}
	\label{optofig11}
\end{figure}

The working principle of the photodetector is similar to that  of the photodiode, transforming photons into an observable current \cite{opto33,opto74} (cf.  FIG. \ref{optofig11}). More specifically, photons transfer energy to electrons, causing them to 'jump'  from the valence band to conduction band (cf.  interband transition). This has a typical timescale of $\sim$1\,ps \cite{opto39}.

\begin{equation}
\gamma({\rm photon}) + e^{-}_{val}\rightarrow e^{-}_{con}\ .
\label{optoeq16}
\end{equation}\\ 
Here, $e^{-}_{val}$ and $e^{-}_{con}$ refer to electrons in the valence and conduction bands respectively.  Bonaccorso et al. \cite{opto33} point out that the absorption bandwidth of light spectra depends upon the choice of semiconductor. As we mentioned before, graphene interacts with an EM range covering the majority of the visible spectrum \cite{opto33,opto39}. Xia et al. \cite{opto75} have also reported that the frequency response of graphene can be upwards of 40\,GHz,  with a theoretical limit reaching even 500\,GHz.   This response generally depends upon the electrical mobility, resistance and capacitance of the materials \cite{opto39}. An appropriate bandwidth for graphene can therefore be adjusted via doping or an applied electric field. Mueller et al. \cite{opto76} further reveal that their results for graphene display a strong photonic response at a wavelength of 1.55\,$\mu$m, when applying the graphene photodetector on fast data communication links.

Bao et al. \cite{opto39} have summarized that current from a photodetector can also be generated just by the contact of graphene and a semiconductor, due to the differing work functions and thermal gradient. Current leakage is one of the major drawbacks of the graphene photodetector, although this can be optimized by reducing the band-gap, or coating some dielectric material on the graphene surface \cite{opto33,opto39}.  Echtermeyer et  al.  \cite{opto77}  show that  a number of metallic nanoparticles can be allocated on the graphene substrate, vastly improving the efficiency of the devices. The basic idea is that a metallic nanoparticle touches the graphene
film, and forms a 'junction like' contact.  Metallic nanoparticles on the graphene layer would thus act as small photodetectors at the same time, and thus enhance the sensitivity \cite{opto77,opto78}. 

\begin{figure}[h]
	\centering
		\includegraphics[width=0.4\textwidth]{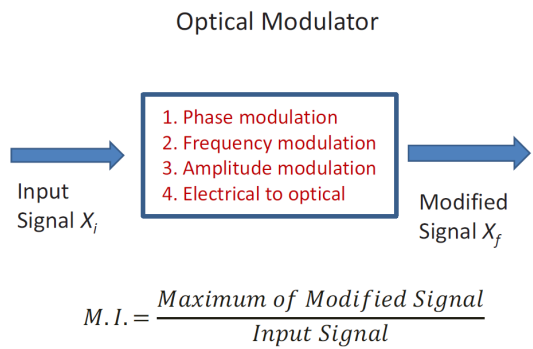}
	\caption{The optical modulator is an important device for converting electrical signals into an optical equivalent, and therefore an ideal bridge between electronic and optical devices. This device can also change the properties of the incident wave, such as the phase, frequency, and amplitude. The Modulation Index $(M.I.)$ is defined as the maximum of the modified signal $X_{f}$, divided by the input signal $X_{i}$.}
	\label{optofig12}
\end{figure}

Some other applications such as the measurement of refractive index \cite{opto79}, and the analysis of metamaterials via the graphene sensor \cite{opto80}, are all being studied on the graphene photonic detector platform.\\ \\
\textbf{B. 	Optical Modulator}\\ \\ 
The optical modulator is a photonic device which transforms electrical signals into an optical equivalent \cite{opto74,opto81,opto82} (cf.   FIG. \ref{optofig12} for a schematic overview).  It  is an essential communication link within many electronic devices, and can also alter the properties of light via doping or an applied electric field \cite{opto39}. For example, assume a plane wave propagates as

\begin{equation}
A = |A| \exp (i\theta)\ ,
\label{optoeq17}
\end{equation}\\
where $A$  can be either electric or magnetic in origin, and $\theta$ is the phase of the wave.  A modulator changes the amplitude $|A|$ and phase $\theta$ of the input wave \cite{opto39}. Graphene is a suitable material for a modulator since it  has a strong response  to a wide range of light spectra (i.e.,  bandwidth) \cite{opto67,opto81,opto82}.  Typically,  graphene will be coated upon the silicon substrate to enhance the absorption rate \cite{opto67}.

Optical modulators can generally be divided into two types \cite{opto39}. The first is an absorptive modulator, converting photons into some other form of energy.   Normally,  an absorption modulator can tune the transmitted light intensity via adjustment of the Fermi energy level $E_{F}$\cite{opto39,opto67}.  The  second type is a refractive modulator which can change the dielectric constant according to variation of the electric field.  Graphene is a promising material for an absorption modulator due to its wide bandwidth and tunable Fermi energy level \cite{opto33,opto39}. Bao et al. \cite{opto39} further  reveal that the interband transition can be tuned to a logical ON/OFF state, dependent upon $E_{F}$ . Regardless, graphene provides a high optical Modulation Index $(M.I.)$, making it an ideal material for any modulator \cite{opto39}. This index is given by 

\begin{equation}
M.I.=\frac{{\rm max}(X_{f})}{X_{i}}\ ,
\label{optoeq18}
\end{equation}\\
where $X_{i}$ and $X_{f}$  refer to the variable before and after modulation respectively.  The graphene modulator can also be applied to the optical resonator, allowing the wavelength to be altered (cf.  for details \cite{opto39}).

Recently, the dielectric sandwich - two layers of graphene with dielectric filling - has been used as an optical signaling modulator \cite{opto81,opto82}.  Gosciniak et al.  \cite{opto81} estimate that  this graphene optical modulator can reach speeds of up to 850\,GHz,  with 3\,dB modulation and small losses. Liu et al. \cite{opto82} have also reported a wide absorption range of 1.35-1.60\,$\mu$m in wavelength.\\ \\ 
\textbf{C. 	Graphene Waveguide}\\ \\ 
A waveguide is a physical channel which traps light, guiding it through a designated path \cite{opto61,opto83}.  For example, fiber-optic cable is a common waveguide for the communication of light signals - its high refractive index $n_{op}$   trapping light inside the fiber \cite{opto61}. As we have already seen, the refractive index depends upon the linear dielectric susceptibility $\chi_{R1}$ \cite{opto39}. Zhang et al.  \cite{opto83}  have studied the wave-modes of the graphene quantum well, identifying energy dispersion relations associated with Klein tunneling and classical wave-modes \cite{opto83}. Zhang et al. \cite{opto83} further note an absence of the third  order classical, and first  order tunneling wave-modes.

\begin{figure}[h]
	\centering
		\includegraphics[width=0.4\textwidth]{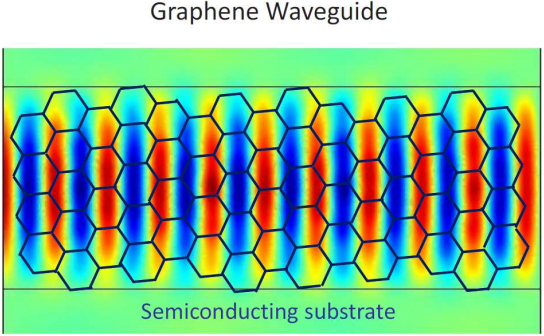}
	\caption{Graphene is coated upon the semiconducting substrate, and either the TE or TM wavemodes can be transmitted along the graphene thin film.}
	\label{optofig13}
\end{figure}

\begin{figure}[h]
	\centering
		\includegraphics[width=0.4\textwidth]{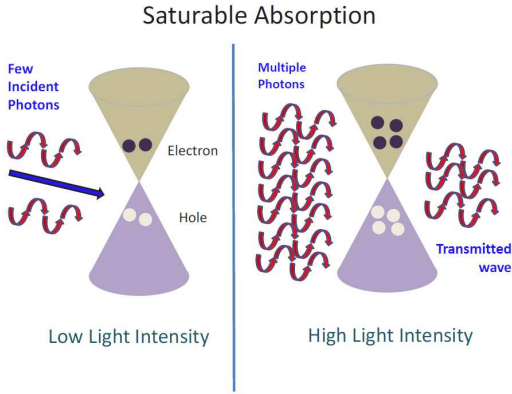}
	\caption{The idea of saturable absorption for graphene is shown. Graphene can absorb photons, and create electron-hole pairs at low incident light intensity. However, electrons are incapable of occupying the conduction band at high light intensity, since most of the states have already been occupied (cf. Pauli exclusion principle).}
	\label{optofig14}
\end{figure}

Graphene plasmonic waveguides have become an essential component for integration with logical devices \cite{opto84,opto85}. Kim et al. \cite{opto84} have studed the plasmonic waveguide for a dielectric substrate coated on graphene, discovering little optical loss and very fast operating speeds. They show that at the peak wavelength $\lambda = 1.31\,\mu$m, the transmission ratio is around 19\,dB for the TM  mode \cite{opto84}.\\ \\ 
\textbf{D. 	Saturable Absorber}\\ \\ 
As  we have already highlighted upon, saturable absorption refers to an absorption of photons decreasing as the light intensity increases \cite{opto66,opto86} (cf.  FIG. \ref{optofig14}).  It is usually applied via the mode-locked laser \cite{opto39,opto87}. Many semiconductors exhibit saturable absorption, but are not as sensitive as graphene \cite{opto39,opto66}.  The  basic idea is as follows - a number of excited electrons occupy the conduction band during high intensity exposure, and electrons in the valence band are no longer able to absorb photons due to the Pauli exclusion principle \cite{opto39,opto87}. This property originates from the nonlinear susceptibility of graphene for a short response time \cite{opto33}. In application, a saturable absorber can be used to transform a continuous wave to a very short wave pulse \cite{opto33}. Generally speaking, monolayer graphene provides a high saturable absorption coeffcient, and recently,  some research has uncovered that CNTs  may also be suitable candidate for a saturable absorber \cite{opto86,opto88}. Bao et al. \cite{opto87} also report that a single layer graphene (SLG) saturable absorber can provide around 66\% modulation depth, and produce picosecond wave pulses.

\ \\ \textbf{E. 	Photonic Limiter}\\ \\ 
A photonic limiter is used to reduce the intensity of light that is emitted from the source \cite{opto33,opto39,opto89,opto90}.  The  mechanism is to permit the passage of low intensity light,  and to filter out light of higher intensity \cite{opto89,opto90}. Dispersed graphene-oxide solutions are generally used for studying the optical limiter \cite{opto86,opto87}. In particular, graphene, has a strong response to a change of light intensity \cite{opto39}, with a transmittance $T({\rm I})$ that  is inversely dependent upon the light intensity I.  Such a device can therefore, for example, be implemented to protect the human eye when working with laser apparatus \cite{opto33}.  Wang  et al.  \cite{opto89} also note how graphene's nonlinear response  is the working principle behind the reduction of light transmitted at high intensity,  and also show that  graphene can limit a wide range of the visible spectrum \cite{opto89,opto90}.

According to Bao  et  al.   \cite{opto39}, the reverse  saturable absorption (opposite to saturable absorption) is the key nonlinear response  that  filters high intensity light,  and subject to certain conditions.  This relates to an optical limiter absorbing more high-energy photons than low-energy photons \cite{opto39}. Lim et al. \cite{opto86} have reported that,  in practice, the property will change from saturable absorption to reverse saturable absorption, only when microplasmas or microbubbles appear. These lead to a nonlinear thermal scattering, which is also an important factor in limiting high intensity light \cite{opto39}. Nevertheless, the graphene photonic limiter is still in the early stages of development, with more drastic efforts being required in the near future.

\section{Transistors}

\begin{figure}[h]
	\centering
		\includegraphics[width=0.4\textwidth]{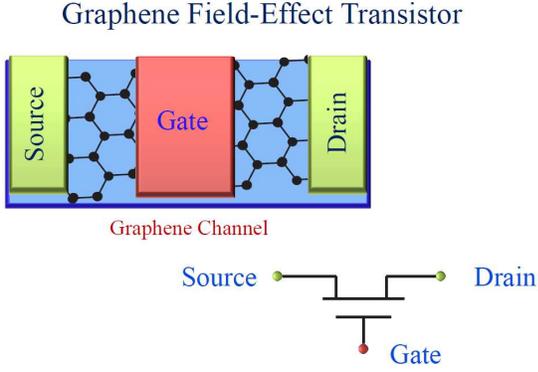}
	\caption{The idea of the graphene FET is shown. The channel of the transistor is made of graphene, and a gate voltage $V_{G}$ controls the current flow $I_{DS}$ from drain to source.}
	\label{optofig15}
\end{figure}

Nowadays, field effect transistors (FETs)  are a key component of most integrated circuitry,  commonly acting as a simple logic gate.   These devices can be of either n-p-n or p-n-p type,  depending upon the desired operation.  In this day and age, when the speed and size of devices are becoming all important, scientists are having to seek revolutionary new materials to replace the likes of Silicon (Si),  Germanium (Ge)  and Gallium  Arsenide (GaAs).   With outstanding electrical mobility, graphene-based materials are becoming evermore prominent as candidates within future transistors (cf. FIG. \ref{optofig15}).
 
\begin{figure}[h]
	\centering
		\includegraphics[width=0.4\textwidth]{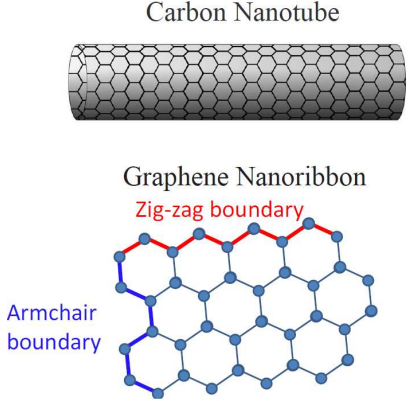}
	\caption{The lower portion  of this  figure  highlights  the  two possible  boundary  conditions  (BCs) that  a  graphene  nanoribbon  (GNR) can  satisfy.  The zig-zag  BC yields  only  a  conducting  state, whereas the armchair  BC can either  imply  a conducting  or semiconducting  state  (dependent  upon the  width  of the  nanoribbon).  The GNR can  also  be curled  to  form  a  carbon  nanotube  (CNT), with a band-gap  that  is inversely proportional  to its  radius.}
	\label{optofig16}
\end{figure}

\ \\ \textbf{A. 	Carbon Nanotube Transistor}\\ \\Before we commence any in-depth discussion of this particular design \cite{opto22,opto23,opto91}, we must first discuss the physical properties of graphene nanoribbons (GNRs)   \cite{opto92}  and carbon nanotubes (CNTs) \cite{opto38,opto93,opto94,opto95,opto96}.  A  GNR is considered  to  be a  piece of graphene of exceptionally narrow width \cite{opto92}. The electrical attributes of GNRs  are determined by their boundary conditions (BCs)  (cf.  FIG. \ref{optofig16}).  The 'armchair' BC  can cause either metallic or semi-metallic behavior to be exhibited, whereas the zig-zag BC  yields only metallic characteristics \cite{opto38,opto93,opto94}. Therefore, GNRs  are another means of generating an energy band-gap. In this case, the gap size is inversely proportional to the nanoribbon width.

Carbon  nanotubes are often considered  to be one-dimensional  structures, and can be formed by curling a GNR (typically  10-100\,nm in width \cite{opto38,opto92}) into a cylindrical configuration (for further details concerning their fabrication, one can refer to \cite{opto38,opto95}).  The nanotubes can either be single-walled or multi-walled, although this must be taken into account when considering the CNT radius $r_{CNT}$ . Since this process leads to structural deformation of graphene's honeycomb lattice, there is an overall modification of the electronic band structure \cite{opto38,opto93,opto94}.  Quantum  equivalents of the capacitance, inductance and resistance have all been exhibited within the electrical properties of CNTs, and an energy band-gap $E_{g}$   is found to be inversely proportional to this CNT radius $r_{CNT}$ \cite{opto38},

\begin{figure}[h]
	\centering
		\includegraphics[width=0.4\textwidth]{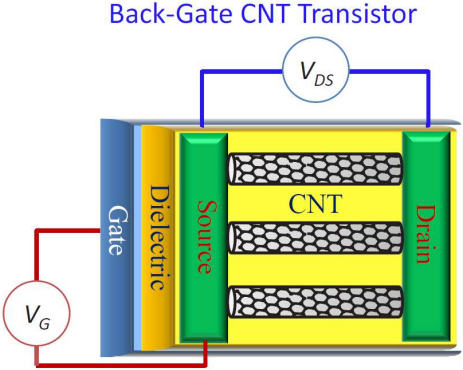}
	\caption{The  figure  shows  the  back-gated   CNT transistor.   It  contains  multiple   CNTs as  the intermediary channels,  providing  stable  performance  and  current  flow. Since the  device  is back-gated,  it can easily be applied  within  integrated  circuitry.}
	\label{optofig17}
\end{figure}

\begin{equation}
E_{g}\sim1/r_{CNT}\ .
\label{optoeq19}
\end{equation}\\ Thus,  together with graphene's capability for long-range ballistic transport (even at room temperature), this presents many useful applications. For example, CNTs  would be an apt source-drain channel within semiconducting devices such as transistors \cite{opto93}.

For decades now, the CNT transistor has been subject to intense study \cite{opto31,opto38,opto93,opto97,opto98,opto99}, with a recently reported high switching ratio \cite{opto99}. They consume much less power, and can possess  shorter channel lengths than their silicon-based counterparts.  They  can also exist in many forms, the most popular being the top-gate, back-gate and wrap-around gate designs \cite{opto93,opto100,opto101,opto102}. Recently,  Shulaker et al. \cite{opto100} have developed a simple computer from the CNT-based  transistors, which can perform more than 20 different instructions. FIG. \ref{optofig17} provides a diagrammatic representation of how a back-gated multi-CNT  transistor may look - the CNTs  themselves acting as the intermediary channels.  Currently, one can produce a purified CNT having less than 0.0001\% impurity - which can minimize any inelastic scattering in the channel \cite{opto101}. The  ON/OFF drain-source current $I_{DS}$  can be tuned by using an  applied electric field (i.e.,  the  gate voltage $V_{G}$)  to  act  upon the  CNT channel \cite{opto101}. Moreover, the ballistic transport of electrons is a result of the one-dimensional CNT structure \cite{opto38}, which again restricts the  degree of inelastic scattering.   CNT transistors would also appear to alleviate the issue of the short-channel effect in silicon-based devices. In theory, the shorter the channel, the faster the transistor \cite{opto31,opto38}. However, usually when the channel has a length scale in tens of nanometers, the drain-source current $I_{DS}$ tends to become most unstable \cite{opto38}. A recent study \cite{opto98,opto101} has revealed that the CNT channel can be as short as 9\,nm, whilst maintaining a stable current.  There are even some predictions that the CNT channel can reach even down to 7\,nm in the near future \cite{opto98,opto101}.

Schottky barriers at the channel-electrode contacts, are the major obstacle with regards to the CNT's application within transistors \cite{opto102}. Specifically, they provide a large resistance at the CNT-electrode interface, due to the differing work-functions \cite{opto38}. The Schottky barrier would generally downgrade the ON/OFF switching ratio \cite{opto38,opto102}. Even worse, this barrier is much larger than for silicon-based devices. A  recent study by Javey  et al. \cite{opto102} reveals that the Schottky barrier would be greatly reduced when using the noble metal, Palladium (Pd) as the electrode. They also show that the CNT channel can even then maintain ballistic transport \cite{opto102}.  It is important to emphasize how both classical and quantum equivalents of inductance, resistance and capacitance are exhibited for CNT transistors \cite{opto38}. In particular, quantum effects become most apparent at  the nanoscale.  Both  quantum inductance and capacitance are determined by the size, BCs  and the density of states (DoS)  of the CNT \cite{opto31,opto38}, whereas the quantum resistance is equal to $h/4e^{2}$.\\ \\{\bf B. 	Tunneling Transistor}\\ \\ {\it 1. 	Mechanisms of Tunneling}\\ \\ As we have already highlighted upon many times now, the absence of a well-defined OFF state in the graphene transistor is a major setback \cite{opto103,opto104}. Assuming a band-gap were to be created, the next hurdle to overcome is the back-current leakage during this OFF state, since this downgrades the power efficiency \cite{opto103,opto104}. Furthermore, opening this band-gap would then reduce the mobility of graphene, with the Dirac fermions being subject to some inelastic scattering \cite{opto104}.

\begin{figure}[h]
	\centering
		\includegraphics[width=0.4\textwidth]{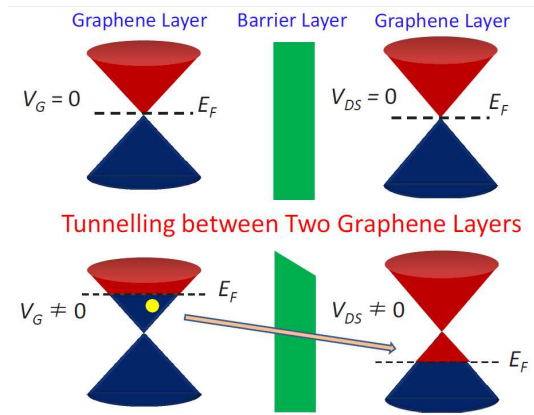}
	\caption{The upper figure shows an intermediate barrier  separating  the two layers  of graphene.  In the  absence  of external electric  fields,  the  Fermi  energies  are  situated  at  the  Dirac points.   When the  gate  voltage  $V_{G}$  and  drain-source   voltages  $V_{DS}$  are  applied,  electrons  begin to accumulate in the conduction  band of one graphene  layer,  and holes in the other.  Tunneling can readily  occur in this  situation, via  a fine tuning  of both voltages.}
	\label{optofig18}
\end{figure}

The tunneling graphene transistor is a revolutionary new concept, and may be capable of alleviating some of the aforementioned  drawbacks.  It  consumes very little  power (up to 109 times less than silicon-based devices \cite{opto105})  and possesses a very fast response time (steep sub-threshold slope) \cite{opto31}.  Michetti  et al.   also report that  an ON/OFF switching ratio can reach as high as $10^{4}$,  even with a small electric field \cite{opto106,opto107,opto108}.   It  is also found that  tunneling occurs at exceptional speed \cite{opto108}. The underlying concept is visualized in FIG. \ref{optofig18}. The interband tunneling is tuned via an applied drain-source voltage $V_{DS}$ and gate voltage $V_{G}$ \cite{opto104,opto109}. Both $V_{DS}$  and $V_{G}$  are used to accumulate the electrons and holes in upper and lower graphene layers respectively,  and thus altering the shape of the potential barrier \cite{opto110}. The tunneling is also associated with the channel length, and thickness of the gate oxide layer $t_{ox}$ \cite{opto106}.
 
\begin{figure}[h]
	\centering
		\includegraphics[width=0.4\textwidth]{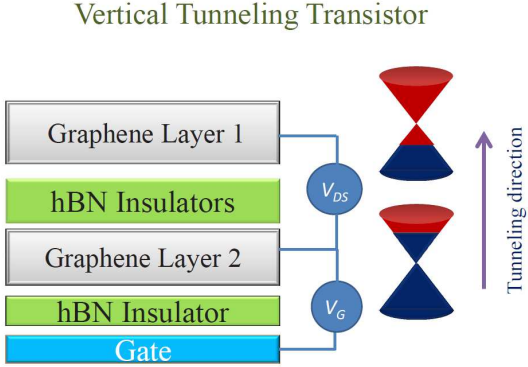}
	\caption{The vertical tunneling  graphene transistor is shown.  The hexagonal  Boron Nitride (hBN) insulating  layers  act  as an intermediary barrier.   The accumulation of holes in the upper graphene layer  is  controlled  by  the  drain-source   voltage  $V_{DS}$,  whereas  the  build-up  electrons  in  the  lower graphene layer  can be tuned via  the gate voltage  $V_{G}$ .  Electrons  are then capable  of tunneling  from the  bottom  to top layer  of graphene.}
	\label{optofig19}
\end{figure}

\ \\ {\it 2. 	Vertical  Design}\\ \\ A  relatively new concept which relies upon vertical  tunneling has been developed by the Manchester research group \cite{opto103,opto104}.  The  graphene-based  device consists of a  few (insulating) layers of hexagonal Boron Nitride (hBN)  or molybdenum disulphide (MoS$_{2}$) \cite{opto111,opto112,opto113}. These are positioned between two graphene sheets which then constitute the electrodes (cf.  FIG. \ref{optofig19}).  The key point here, is that the insulating layers act as a barrier, and thus preventing the flow of current. As such, there is no need for a well-defined band-gap in graphene \cite{opto111,opto112,opto113}. This  has the added benefit of greatly reducing any  current leakage whilst in the OFF state \cite{opto111,opto112,opto113}.  The whole process of current tunneling then acts perpendicular to the layers \cite{opto111,opto112,opto113}.  Electrons in the bottom graphene layer will begin to accumulate once a gate voltage $V_{G}$  is applied across the lower insulating layer \cite{opto114,opto115}. The  drain-source voltage $V_{DS}$  is then added to create holes in the upper graphene layer \cite{opto103,opto111,opto112,opto113}. This  has the desired effect of increasing the Fermi energy $E_{F}$    in bottom graphene layer, and decreasing $E_{F}$    in the upper layer.  Electrons in the bottom graphene layer are then capable of tunneling to  the top graphene layer \cite{opto104}.  A  recent study by Georgiou et  al. \cite{opto111}  reveals that  the current modulation can reach a high value of $10^{6}$ (even at room temperature). It is also interesting to note that a resonant tunneling within the vertical transistor occurs in some energy states, and a negative differential conductance exists (i.e.,  current decreases upon an increase in voltage) \cite{opto112,opto113}.

\ \\ {\bf C. 	High  Frequency Devices}\\ \\ High frequency transistors do not require an OFF state, and can operate solely through variations of the current or voltage signaling \cite{opto31,opto116}. Graphene may thus be applicable within the realms of high-frequency transistors, inverters, or operational amplifiers \cite{opto31,opto84,opto117}. Graphene's response to these signals is incredibly fast, with operating speeds of around a few hundred GHz  \cite{opto84,opto116,opto118,opto119,opto120}.

The performance of high frequency devices is characterized by two important parameters - the cut-off frequency $f_{cut}$  and maximum oscillation frequency $f_{max}$. The cut-off frequency $f_{cut}$  is given by a current gain $G_{I}$  equal to unity \cite{opto116}

\begin{equation}
G_{I}=20\log_{10}\left(\frac{I_{out}}{I_{in}}\right)\ ,
\label{optoeq20}
\end{equation}\\ where $I_{out}$ and $I_{in}$ are the output and input currents respectively.  Typically, $f_{cut}$  is proportional to the trans-conductance $g_{rf}$   and the thickness of the gate oxide layer $t_{ox}$, whereas inversely proportional to the transistor gate length $L_{G}$   and gate width $W_{G}$ \cite{opto31,opto116}. The whole expression is given by

\begin{equation}
f_{cut}=c_{1}\,\frac{t_{ox}\,g_{rf}}{L_{G}\,W_{G}}\ ,          
\label{optoeq21}
\end{equation}\\ where $c_{1}$ is a constant associated with dielectric gate.  In experiments, one would only shorten the gate length $L_{G}$   for simplicity,  thereby increasing the cut-off frequency.   Wu et al \cite{opto116} report that with CVD-prepared graphene, $f_{cut}$  can reach upwards of 155\,GHz  for a relatively short gate length of 40\,nm.  Theoretical simulations have indicated that a cut-off frequency of 1\,THz  can be attained for just a few nanometers gate length \cite{opto118}.

Similar to $f_{cut}$, the maximum oscillation frequency $f_{max}$  is obtained for a power gain $G_{P}$ equal to one.  Here, we have $G_{P}=10\log_{10}(P_{out}/P_{in})$,  where $P_{out}$ and $P_{in}$  are the output and input powers respectively \cite{opto116}. The value of $f_{max}$  in graphene-based devices is slightly more complicated, and is dependent upon the cut-off frequency,  gate resistances, and the trans-conductance $g_{rf}$ \cite{opto116,opto118}. A recent report \cite{opto116} has mentioned that $f_{max}$  can reach up to 20\,GHz.   However, it is important to note that a short gate length would not necessarily imply a  high value for $f_{max}$ \cite{opto116,opto118}.  At  present, not much is understood of the $I-V$ characteristic curve - this has three regions, firstly linear, then saturating,  and finally a second linear region \cite{opto31,opto116,opto118}.  In addition, a change of gate voltage $V_{G}$  would alter the shape of the $I-V$ curve, and making the saturation region ambiguous. Without a stable saturation region, the value of $f_{max}$  is limited.  This problem will require urgent attention in the near future, if the high-frequency transistor is to make any headway \cite{opto84,opto121}.

\section{Summary}

Graphene's  outstanding capabilities have drawn the  attention  of  scientists from several interdisciplinary backgrounds - all looking to take advantage.   This  stand-alone two-dimensional structure is a playground for Dirac fermions which possess a zero effective mass \cite{opto4,opto122,opto123}.  Quantum  phenomena have been observed even at room temperature; a series of anomalous quantum effects including QHE  and Klein tunneling \cite{opto1,opto54}. Graphene's versatility is nigh on endless - in this paper, we have merely focused upon optoelectronic devices and transistors.  Optical  communications provide a much wider bandwidth, with higher efficiencies  than most typical  conducting wire.  We are thus dawning upon a new golden photonic age of higher internet speeds, and entertainment-based devices. Graphene's high transmittance, strong interaction of light with ultra-fast response time, wide absorption spectrum, and tunable optical conductivity \cite{opto33,opto39,opto67}, present an ideal material for optical devices! Amongst many others, these include the photodetector, optical modulator, plasmonic waveguide and also the saturable absorber. The absence of any discernible band-gap for graphene is an unavoidable issue for logical devices, although one may be created via various methods (e.g.,  structural deformation or chemical doping) \cite{opto31}. The CNT transistor is now a well-established  technology - developed over more than 30 years. Only now has the dream of a CNT-based computer become a working reality \cite{opto100}.  Carriers in CNT channels can perform ballistic transport, even for very short lengths.   However, the graphene vertical tunneling transistor is something rather novel. This device itself does not require band-gap at all,  and yet,  both operates at exceptional speeds, whilst consuming very little power. Our final talking point was the high frequency transistor, which acts as amplifier in the circuit rather than a typical logical device.  The cut-off frequency can reach theoretical estimates of up to 1\,THz,  for just a few nanometers of gate length \cite{opto38}. Although  we have plainly made the case for graphene's implementation within numerous optical and electronic devices, there are a few obstacles which we must overcome. These are the nonlinear $I-V$  characteristic curve, and the emergence of Schottky  barriers (although we mention a suitable fix).  Fifty  years ago, no one would have ever envisaged that optical or silicon-based devices would have their place in everyday life.  Graphene may change the world!

\end{document}